\documentclass[pre,aps,nofootinbib,twocolumn,floatfix,superscriptaddress]{revtex4}

\usepackage{epstopdf}
\usepackage{subfigure}
\usepackage{amsmath}
\usepackage{amssymb}
\usepackage{amsthm}
\usepackage{amscd}
\usepackage[usenames]{color}
\usepackage{graphicx}

\def\th{\theta}

\def\del{\nabla}

\def\N{\textbf{N}}

\def\sn{\textrm{sn}}
\def\cn{\textrm{cn}}
\def\dn{\textrm{dn}}
\def\ns{\textrm{ns}}

\def\cs{\textrm{cs}}
\def\sc{\textrm{sc}}
\def\ds{\textrm{ds}}

\def\am{\textrm{am}}

\def\Im{\textrm{Im}}

\def\Rs{R^2}

\begin{document}

\title{Smectic Pores and Defect Cores}

\author{Elisabetta A. Matsumoto}
\email{sabetta.matsumoto@gmail.com}
\affiliation{Princeton Center for Theoretical Science, Princeton University, Princeton, NJ 08544, U.S.A.}
\affiliation{Department of Physics \& Astronomy, University of Pennsylvania, Philadelphia PA 19104, U.S.A.}
\author{Randall D. Kamien}
\affiliation{Department of Physics \& Astronomy, University of Pennsylvania, Philadelphia PA 19104, U.S.A.}
\author{Christian D. Santangelo}
\affiliation{Department of Physics, University of Massachusetts, Amherst, MA 01003, U.S.A.}

\begin{abstract}
Riemann's minimal surfaces are a complete, embeddable, one-parameter family of minimal surfaces with translational symmetry along one direction.  It's infinite number of planar ends are joined together by an array of necks, closely matching the morphology of a bicontinuous, lamellar system with pores connecting alternating layers.  We demonstrate explicitly that Riemann's minimal surfaces are composed of a nonlinear sum of two oppositely-handed helicoids.  This description is particularly appropriate for describing smectic liquid crystals containing two screw dislocations.
\end{abstract}

\maketitle

\section{Introduction}

The connection between statistical mechanics and the calculus of variations is both powerful and profound.  Because geometry and elasticity intimately comingle in soft materials, the resulting equations of equilibrium are often described or approximated by purely geometrical considerations; interfaces, in particular, classically adopt the morphology of constant mean curvature surfaces, typically minimal surfaces.  The classic soap film problem boils down to minimizing surface area subject to prescribed boundary conditions.  Smectic liquid crystals introduce an additional layer of complexity to this classic problem.  Although made of layers, smectics develop periodic order normal to the layers, while each layer remains fluid-like.

Isolating the topological excitations proves extremely powerful in the study of ordered media.  Many key physical properties may be described entirely by their topological defects.   Indeed, in the two-dimensional XY model, it is possible to study the energetics and phase transitions entirely in terms of topological variables.  Likewise, topological defects form in smectics and, in some descriptions \cite{Nelson:1981p363, Toner:1982p462}, are the principal players in the melting of the smectic phase into the nematic phase.

Moreover, the effects of surface curvature and topological defects are inextricably linked.  While it is simple to see how a single topological defect in a surface or family of surfaces induces curvature into its host, the converse problem is not nearly as transparent.  Similarly, the addition of further defects into an already curved surface is not merely a superposition, nor are the interactions of a finite number of defects linear.  However, the topology of the surfaces themselves provides a framework for studying the collective effects of curvature and topological defects.  A deep theorem in differential geometry states that all minimal surfaces are locally constructed out of pieces of helicoids and catenoids \cite{Colding:2006p11106}.

An explicit deconstruction of surfaces into their topological constituents provides a unique and visual language with which to describe the properties of many diverse systems.   In the case of smectics, we have found it useful to construct smectic textures by viewing the layers as the Riemann surface of meromorphic function on the plane encoding the two-dimensional arrangement of three-dimensional line defects \cite{Kamien:2001p797, Santangelo:2006p137801, Matsumoto:2009p257804}. The one-dimensional periodicity arises   from the multiple sheets formed by logarithmic branch points.  Again, great simplification occurs in those configurations that can be represented in this way:  the gradient terms of the free energy specify harmonic solutions, which are highly amenable to this Riemann surface description.

Lattices of passages and pores often appear in both membranes and bicontinuous systems \cite{Carvalho:1994p3321,Charitat:1997p373}.  In smectics, this morphology is created by a pair of oppositely-handed screw dislocations \cite{Bouligand:1972p525}.  Because Riemann's minimal surfaces are constructed from a pair of oppositely-handed helicoids \cite{Colding:2006p11106}, it shall serve as our model for a system of pores in a smectic.  In order to calculate the stability of this smectic phase, the exact locations of the screw dislocations must be known.  The morphology of the dislocation cores will also allow us to study the crossover between locally helicoidal morphology and a network of pores.

In the following we will review the free energy functional of the smectic liquid crystal, introduce our construction for including topological defects in a phase field construction, and discuss a new manifest decomposition of Riemann's minimal surfaces into a pair of helicoids.

\section{Smectic Free Energy, Smectic Topology, and Minimal Surfaces}

As smectics are a one-dimensional crystal of two-dimensional layers of liquid with equilibrium spacing $d=2 \pi/q_{\textrm{sm}}$, it is apt to describe them with a complex order parameter $\psi(\textbf{x})=\vert \psi_0(\textbf{x}) \vert \exp \left(-i q_{\textrm{sm}} \Phi(\textbf{x})\right)$.  The magnitude of the order parameter, $\vert \psi_0(\textbf{x}) \vert,$ determines whether or not smectic order is present.  However, all subsequent calculations occur deep within the smectic phase, where $\psi_0(\textbf{x})=\psi_0$ is constant away from defect cores which we treat as cylindrical punctures.  The phase of the smectic order parameter defines the smectic layers; this may be seen by considering the mass-density wave representation of the smectic, where layers are defined by level sets of the phase field $\Phi(\textbf{x})=n d, \, n\in\mathbb{Z}$.

In addition to the standard membrane curvature energy terms due to the fluid nature of the smectic layers, smectics are also penalized if the layer spacing deviates from the equilibrium value.  Therefore, the conventional form of the rotationally invariant, nonlinear smectic free energy is given by
\begin{equation}
F_{\rm sm}=B \int d^3x \left\{\left[\left(\nabla \Phi\right)^2-1\right]^2+\lambda^2\left(\nabla\cdot \textbf{N}\right)^2\right\},
\end{equation}
where $B$ is the bulk modulus, $\lambda$ is the penetration depth for splay, and ${\bf N}=\nabla\Phi/\vert \nabla \Phi \vert$ is the unit normal vector for the layers \cite{Kamien:2006p229,Chaikin:1995}.  An additional term due to the Gaussian curvature may be neglected due to the Gauss-Bonnet theorem.

In smectics the simplest case of a topological defect, the screw dislocation, joins together an infinite number of flat layers along a straight line, taking the form of a helicoid, $\Phi=z-\tan^{-1}\left(y/x\right)=z-\Im\bigl[\log(x+iy)\bigr]$.  The compression energy diverges along the line $\{x=0,y=0,z\in\mathbb{R}\}$ \cite{deGennes:1993,Chaikin:1995}.  To prevent this divergent contribution, the smectic locally melts along the defect line, saving us from a topological conundrum at the cost of introducing
a new set of boundary conditions into the problem.  While exact solutions to the Euler-Lagrange equations often prove elusive, surfaces approximating the real layer structure are easily obtained, thanks to the zoo of minimal surfaces.  Choosing a minimal surface with the right topology as our starting point, the locations where the compression energy diverges are known and we can identify the defects.  Using one-parameter families of minimal surfaces as variational solutions gives us a handle on the energetics of these systems.

Merely specifying the zeroes and poles in a phase field completely defines the topology of the resulting surface.  For regular arrangements of defects in two-dimensions, elliptic functions are the building blocks of the textures.  The properties of elliptic functions vastly simplify both the analytic and numeric calculations of the energy.  A simple example is afforded by Scherk's first surface \cite{Scherk:1835p185}, with the multivalued height function
\begin{eqnarray}
h(x,y) &=& -\sec(\frac{1}{2}\alpha)\tan^{-1}\left\{\frac{\tanh[\frac{1}{2}x\sin\alpha]}{\tan[y\sin(\frac{1}{2}\alpha)]}\right\}\nonumber\\
 &=&- \sec(\frac{1}{2}\alpha)\arg \sin\left[y +i x\cos(\frac{\alpha}{2})\right].
 \end{eqnarray}
The last equality expresses the height function as the argument of a trigonometric function \cite{Kamien:2001p797}.

Because Riemann's minimal surfaces have an infinite number of planar ends, their boundary conditions and asymptotic behaviour make them extremely appealing to model the interaction between defects in a smectic.  In the following, we will provide a precise decomposition of the one-parameter family of Riemann's minimal surfaces into a pair of oppositely charged helicoids.  Like the decomposition of Scherk's first surface into defects \cite{Kamien:2001p797}, this allows us to consider the defects as the principle players. The particular closed form expression for Riemann's minimal surfaces reveals an explicit parametrization the cores of the defects.  Enforcing the smectic boundary condition-- flat, equally spaced layers at infinity -- while maintaining the topological defects necessitates curvature in the dislocation cores.

\section{Riemann's minimal surfaces}

Riemann's minimal surfaces are a complete, embedded one-parameter family of minimal surfaces \cite{Riemann:1898,Lopez:1997p376}.  Riemann's minimal surfaces are foliated by circles and, thus, may be described by a phase field of the following form:
\begin{equation}
\Phi=\big[x-\alpha(z)\big]^2+\big[y-\beta(z)\big]^2-R^2(z).
\end{equation}\label{ph1}
A minimal surface, by definition, must satisfy the condition: $\rm{H}=\frac{1}{2} \, \del \cdot {\bf N}=0$ for the unit layer normal $\mathbf{N}$.

\begin{figure*}[t!]
\centerline{
\includegraphics[width=84mm]{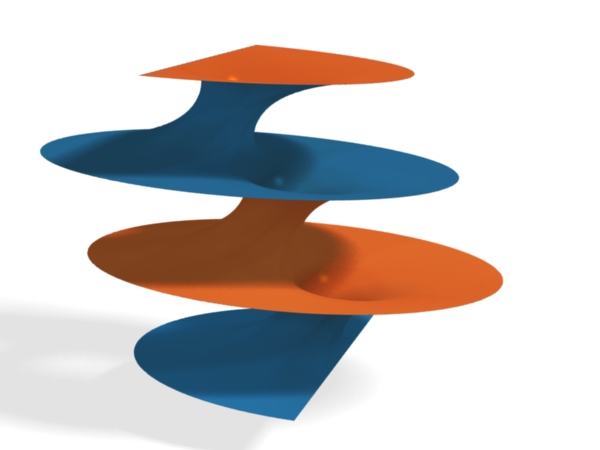}
\includegraphics[width=84mm]{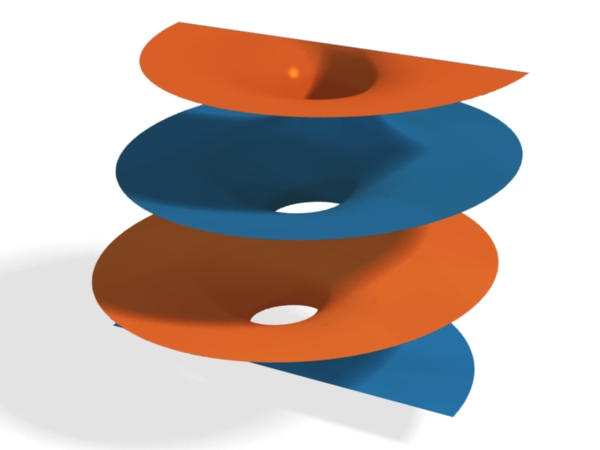}
}
\caption{\label{fig1}
Riemann's minimal surfaces are a complete, one parameter family of minimal surfaces foliated by circles.  Due to their morphology, they are models for a bicontinuous phase in which a network of pores connects neighbouring regions.  In particular, Riemann's minimal surface has an infinite number of planar ends, which makes them particularly appealing as a model for smectic liquid crystals.
}
\end{figure*}

\subsection{Weierstrass elliptic functions}
Following Nitsche \cite{Nitsche:1989}, the phase field must satisfy:
\begin{equation}\label{de1}
\bigg[ \frac{\Phi_z}{R^2(z)} \bigg]_z + \frac{2}{R^2(z)} =0,
\end{equation}
from which it follows that $\Rs(z)$ satisfies
\begin{equation}\label{Rs'}
\Big\{\big[\Rs(z)\big]'\Big\}^2 = 4\Big[d^2\, R^6(z)+2\, c \, R^4(z)-R^2(z) \Big]
\end{equation}
with $d^2=a^2+b^2$.  Similarly, $\alpha'(z) = a \, \Rs(z)$ and $\beta'(z) = b \, \Rs(z)$.  
By considering solutions of the form $\Rs(z)=A \, f(z) + B,$ equation (\ref{Rs'}) becomes the familiar equation for the Weierstrass $\wp$-function, $\big[\wp'(z) \big]^2=4 \, \wp^3(z) -g_2 \, \wp(z) -g_3 = 4 \, \big[\wp(z)-e_1\big]\big[\wp(z)-e_2\big]\big[\wp(z)-e_3\big],$ where $e_1+e_2+e_3=0.$  We obtain the following solution for $\Rs(z):$
\begin{equation}\label{Rs}
\Rs(z)=\frac{\wp\big(z,\{g_2,g_3\}\big) - 2 \, c/3}{a^2+b^2},
\end{equation}
where $g_2=4(a^2+b^2+\frac{4}{3}c^2)$ and $g_3=-\frac{8}{3} \, c \, (a^2+b^2+\frac{8}{9}c^2)$ are the Weierstrass invariants.  For reference, the half-period values are
\begin{eqnarray}\label{zeroes}
e_1 &=& \wp(\omega_1 /2)=-\frac{c}{3}+\sqrt{a^2+b^2+c^2}\nonumber\\
e_2 &=& \wp(\omega_1 /2+\omega_3 /2)=\frac{2}{3} \, c\nonumber\\
e_3 &=& \wp(\omega_3 /2)=-\frac{c}{3}-\sqrt{a^2+b^2+c^2},
\end{eqnarray}
for $c<0.$  Without loss of generality, we set $a=0,$ as this merely fixes the free rotations about the $z-$axis.

\subsection{Jacobi elliptic functions}
It behooves us to work with Jacobi elliptic functions for the remainder of this paper.  Jacobi elliptic functions are doubly periodic with real period $2 \, K(m)$ and imaginary period $4 \, i \, K(1-m),$ where $m$ is the square of the elliptic modulus and $K(m) = \int_{0}^{\pi/2} dt/\sqrt{1-m \, \sin^2(t)}$ is the complete elliptic integral of the first kind.  To convert between the two equivalent representations of elliptic functions, Weierstrass and Jacobi, we use the identity:
\begin{equation}\label{weierstrass2jacobi}
\wp(z) =e_3 + (e_1-e_3) \, \ns^2 (z \, \sqrt{e_1-e_3},m),
\end{equation}
where $m=(e_2-e_3)/(e_1-e_3)$ 
\footnote{We use the standard Glaisher convention for notation ${\rm pq}(z,m)=\displaystyle \frac{{\rm pr}(z,m)}{{\rm qr}(z,m)},$ where p, q, and r can be any of the following s, c, d, or n and ${\rm ss}={\rm cc}={\rm dd}={\rm nn}=1,$ \cite{Abramowitz:1964,Walker:1996}.  For example, $\cs(z,m)=\displaystyle \frac{\cn(z,m)}{\sn(z,m)}.$}.  
Therefore, in terms of Jacobi functions,
\begin{widetext}
\begin{equation}
R^2(z) = \frac{e_3-e_2+(e_1-e_3)\ns^2(z\sqrt{e_1-e_3},m)}{b^2}=\frac{e_1-e_3}{b^2}\big(\ns^2(\sigma z,m)-m\big)=\frac{\sigma^2}{b^2}\ds^2(\sigma z,m),
\end{equation}
where $\sigma^2=e_1-e_3=2\sqrt{b^2-c^2}=b/\sqrt{m(1-m)},$ and we have used the identity $\dn^2(z,m)+m\, \sn^2(z,m)=1,$. The integral defining the center of the circles is then
\begin{equation}
\int dz \, \ds^2(z,m) = z(1-m) -\frac{ \cn(z,m)  \dn(z,m)}{\sn(z,m)} - E \big[ \am(z,m),m \big],
\end{equation}
where $E(z,m)=\int_0^z dt\sqrt{1-m \sin^2z}$ is the incomplete elliptic integral of the second kind. Note that the integral $E\big[\am(z,m),m\big]$ is composed of a linear part, $z \, E(m)/K(m),$ and a periodic part, $Z \big[\am(z,m),m\big]$. In terms of the Jacobi elliptic functions, the phase field is
\begin{equation}\label{jacobi}
\Phi= x^2 + \bigg\{y -\frac{\sigma}{b}\Big[ \tilde{z}(1-m)- E\big( \am(\tilde z,m),m\big) + \cs(\tilde{z},m) \dn(\tilde{z},m)\Big] \bigg\}^2  -\frac{\sigma^2}{b^2}  \ds^2 \big(\tilde{z},m\big)=0,
\end{equation}
\end{widetext}
where $\tilde{z}=\sigma z.$  Since $\sigma z$ is merely a rescaling of the $z$-axis, we have the freedom to choose the constant to be $\sigma=2 K(m)/d$ to ensure that the periodicity of the phase field matches the ideal spacing of the smectic layers $d$. The final parameter, $b=4 K(m)^2\sqrt{m(1-m)}$, may now be expressed in terms of the elliptic modulus.  The resulting surface is now a one-parameter family of surfaces, depending on $m \in [0,1].$ 

As there are an infinite number of phase fields that describe the same surface, determining the appropriate representation of the phase field for each given application requires careful consideration.  First, we demonstrate that Riemann's minimal surfaces are composed topologically of two screw dislocations of opposite handedness, and, secondly, we find a form of the phase field for Riemann's minimal surfaces that satisfies the boundary conditions of a smectic liquid crystal.

Expanding the square in $y$ in the phase field, we obtain
\begin{eqnarray}\label{jacobi2}
\Phi &=& x^2 + \big[y-\zeta(\tilde{z},m)\big]^2-\eta^2(\tilde{z},m)\nonumber\\
&+&2 \frac{\sigma}{b} \big[y-\zeta(\tilde{z},m)\big] \, \cs(\tilde{z},m) \, \dn(\tilde{z},m)=0,
\end{eqnarray}
where we define $\zeta(z,m) = (\sigma/b)\big[z(1-m)- E\big(\am(z,m),m\big)\big],$ and $\eta^2 (z,m) = \frac{\sigma^2}{b^2}\big[\ds^2(z,m)-\cs^2(z,m)\dn^2(z,m)\big]=(\sigma^2/b^2) \dn^2(z,m).$  After some algebra, we find the equivalent phase field,
\begin{widetext}
\begin{equation}\label{jacobi3}
\Phi = 1 + \frac{\big[y-\zeta(\tilde{z},m)\big]^2}{x^2 - \eta^2(\tilde{z},m)} - \bigg[\frac{y-\zeta(\tilde{z},m)}{x+\eta(\tilde{z},m)}-\frac{y-\zeta(\tilde{z},m)}{x-\eta(\tilde{z},m)} \bigg] \frac{\sigma\cs(\tilde{z},m) \dn(\tilde{z},m)}{b\, \eta(\tilde{z},m)}  = 0.
\end{equation}
\end{widetext}
This provides a relation between $y$, $\tilde z$,  and $x$,
\begin{equation}\label{jacobi4}
\frac{\frac{y-\zeta(\tilde{z},m)}{x + \eta(\tilde{z},m)} - \frac{y-\zeta(\tilde{z},m)}{x - \eta(\tilde{z},m)}}{1+ \frac{y-\zeta(\tilde{z},m)}{x+\eta(\tilde{z},m)} \, \frac{y-\zeta(\tilde{z},m)}{x-\eta(\tilde{z},m)}} = \sc(\tilde{z},m),
\end{equation}
where we note that the definition of $\sc(z,m)=\sin\big(\am(z,m)\big)/\cos\big(\am(z,m)\big).$  Taking the arctangent of both sides and rearranging the terms yields the phase field,
\begin{eqnarray}\label{jacobi5}
\Phi&=&\am(\sigma z,m) + \tan^{-1} \bigg[ \frac{y-\zeta(\sigma z,m)}{x-\eta(\sigma z,m)} \bigg]\nonumber\\
&-&\tan^{-1} \bigg[ \frac{y-\zeta(\sigma z,m)}{x + \eta(\sigma z,m)} \bigg]= 0,
\end{eqnarray}
which is clearly a nonlinear sum of two helicoids of opposite handedness.  The cores of the defects are located at $x=\pm\eta(\sigma z,m), \ y=\zeta(\sigma z,m)$, where $\zeta(\sigma z,m)$ consists of a linear component with a periodic term superimposed upon it, and $\eta(\sigma z,m)$ is a periodic function with maximum at $\eta(0,m)=\frac{\sigma}{b}$ and minimum at $\eta(K(m),m)=\frac{\sigma}{b}\sqrt{1-m}.$

In order to satisfy the smectic boundary condition, flat evenly spaced layers infinitely far from the defects, the phase field must be normalized such that the compression energy vanishes at infinity, \textit{i.e.} $\displaystyle \lim_{x,y \to \infty}(\nabla \Phi)^2=1$. Since the derivative $\partial_{z}\am(z ,m)=\dn(z,m)$ is not independent of $z$, this phase field does not satisfy this boundary condition. One final manipulation, based on the identity $F\big(\am(z,m),m\big)=z,$ where $F(z,m)=\int_0^zdt(1-m\sin^2t)^{-1/2}$ is the incomplete elliptic integral of the first kind, gives the phase field representation of Riemann's minimal surfaces,
\begin{widetext}
\begin{equation}\label{phasefield}
\Phi=z+\frac{1}{\sigma}F\Bigg[ \tan^{-1} \bigg( \frac{y-\zeta(\sigma z,m)}{x-\eta(\sigma z,m)} \bigg)  -\tan^{-1} \bigg( \frac{y-\zeta(\sigma z,m)}{x + \eta(\sigma z,m)} \bigg),m\Bigg]= 0.
\end{equation}
\end{widetext}
This is our main result.

\section{Pores and Defect Cores}

\begin{figure}[h!]
\includegraphics[width=60mm]{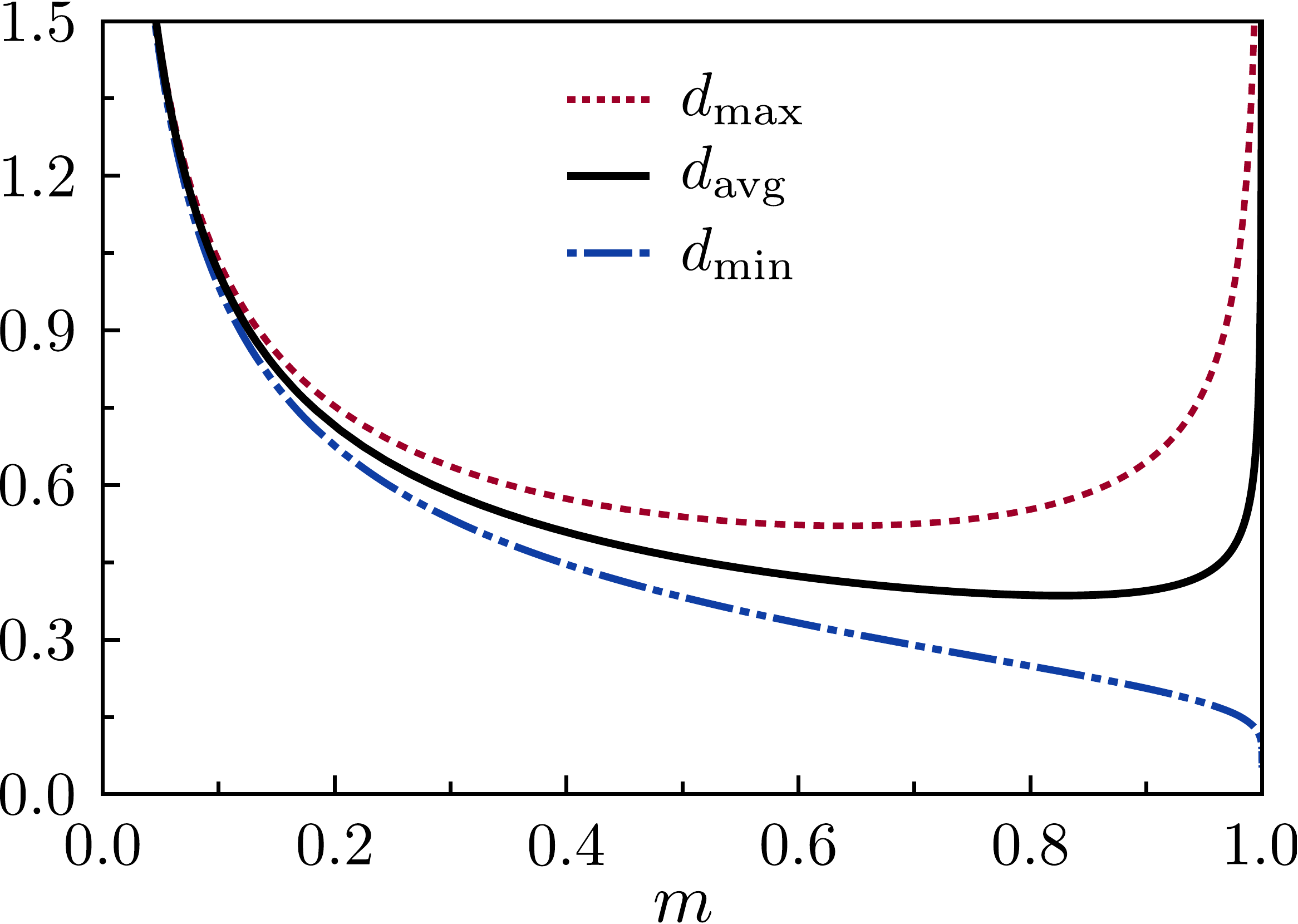}
\caption{\label{fig:separation}
Due to the curvature of the defect cores, the elliptic modulus parametrizes the distance between the two helicoids.  The maximum, minimum, and average separation are given by $d_{\rm max}=\frac{1}{\sigma\sqrt{m}},$ $d_{\rm min}=\frac{1}{\sigma\sqrt{m(1-m)}},$ and $d_{\rm avg}=\frac{\pi}{\sigma^2\sqrt{m(1-m)}},$ respectively.
}
\end{figure}

The entire one-parameter family of Riemann's minimal surfaces are parametrized by the elliptic modulus, $m \in [0,1]$. As it sweeps through all allowed values, the morphology of the surface, and simultaneously the shape of the defects, changes.  In the limit that $m\rightarrow 0,$ the defect cores are straight and infinitely far apart.  However, as $m$ increases, the distance of closest approach between the two defect cores decreases. 
The elliptic modulus is a proxy for the distance between the two helicoids, where the maximum, minimum and average separation of the defects are respectively $d_{\rm max}=\frac{1}{\sigma\sqrt{m}},$ $d_{\rm min}=\frac{1}{\sigma\sqrt{m(1-m)}},$ and $d_{\rm avg}=\frac{\pi}{\sigma^2\sqrt{m(1-m)}},$ shown in Fig.~\ref{fig:separation}.
If the layers are to remain flat along the boundary at infinity, the cores necessarily bend.  Conversely, two straight screw dislocations in a smectic will force the layers to bend so that $H \ne 0$.  The amplitude of the distortion of the cores also must increase as $m$ increases. Nevertheless, the topological character of straight screw dislocations is preserved even though the cores of the dislocations possess curvature.  The line integral of the director field around a closed loop encircling a defect, $\oint\N\cdot d{\bf l}=\pm d,$ remains constant regardless of the position or orientation of the core.

\begin{figure}[h!]
\raisebox{0.25\linewidth}{(a)}
\includegraphics[width=76mm]{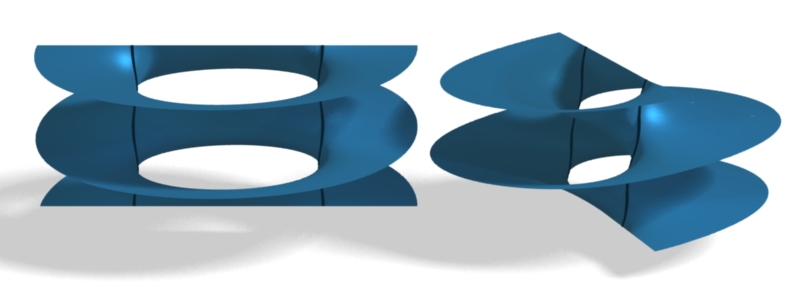}
\raisebox{0.25\linewidth}{(b)}
\includegraphics[width=76mm]{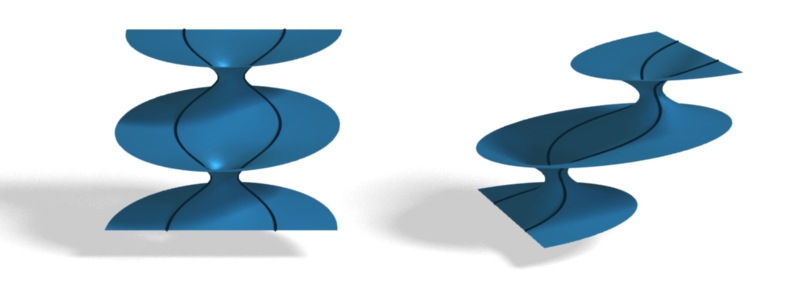}
\caption{\label{fig:elliptic_modulus}
Two of Riemann's minimal surfaces are shown here for parameter values (a) $m=0.075$ and (b) $m=0.925$.
}
\end{figure}

Alternatively, we might have used the construction $\Phi$ via foliation of circles implies that layers flatten when the radius approaches infinity and that adjoining layers are connected by necks.  By varying the surface parameter through the range $m\in[0,1],$ the family of Riemann's minimal surfaces transforms between these two morphologies, as seen in FIG.~\ref{fig:elliptic_modulus}.

\section{Energetics}

Comparing the energetic cost of Riemann's minimal surfaces to that of other smectic textures determines their relative stabilities.  To illustrate how we do this, we first consider the case of a single screw dislocation.
For such a comparison to be valid, all of the divergences must be treated identically.  While it is commonplace to cut off the integration in cylindrical regions, where varying the radius determines the functional form of the divergence, the curved nature of the defects precluded straightforward implementation of this method.  Instead, we opt for a more physically-motivated cutoff: the defect core is identified with the volume in which the energy density exceeds a critical value $e_0.$  This cutoff captures the fact that the real smectic melts when the local energy density exceeds the condensation energy for forming a smectic.
The single dislocation, $\Phi_{\rm screw}=z-(1/\pi) \tan^{-1}\big(y/x\big),$ has compression energy density $f_{\rm screw}=
\big[(\nabla \Phi_{\rm screw})^2-1\big]^2=1/\left[\pi^4 \left(x^2+y^2\right)^2\right]=\big(\pi r)^{-4}.$  The energy density exceeds $e_0$ for $r<r_0=1/(\pi e_0^{1/4}),$ corresponding to a free energy per unit length of $F_{\rm screw}=1/(\pi^4) \int_0^{2\pi}d\th\int_{r_0}^\infty r dr r^{-4}=\sqrt{e_0}/\pi.$

\begin{figure}[ht]
\centerline{
\includegraphics[width=84mm]{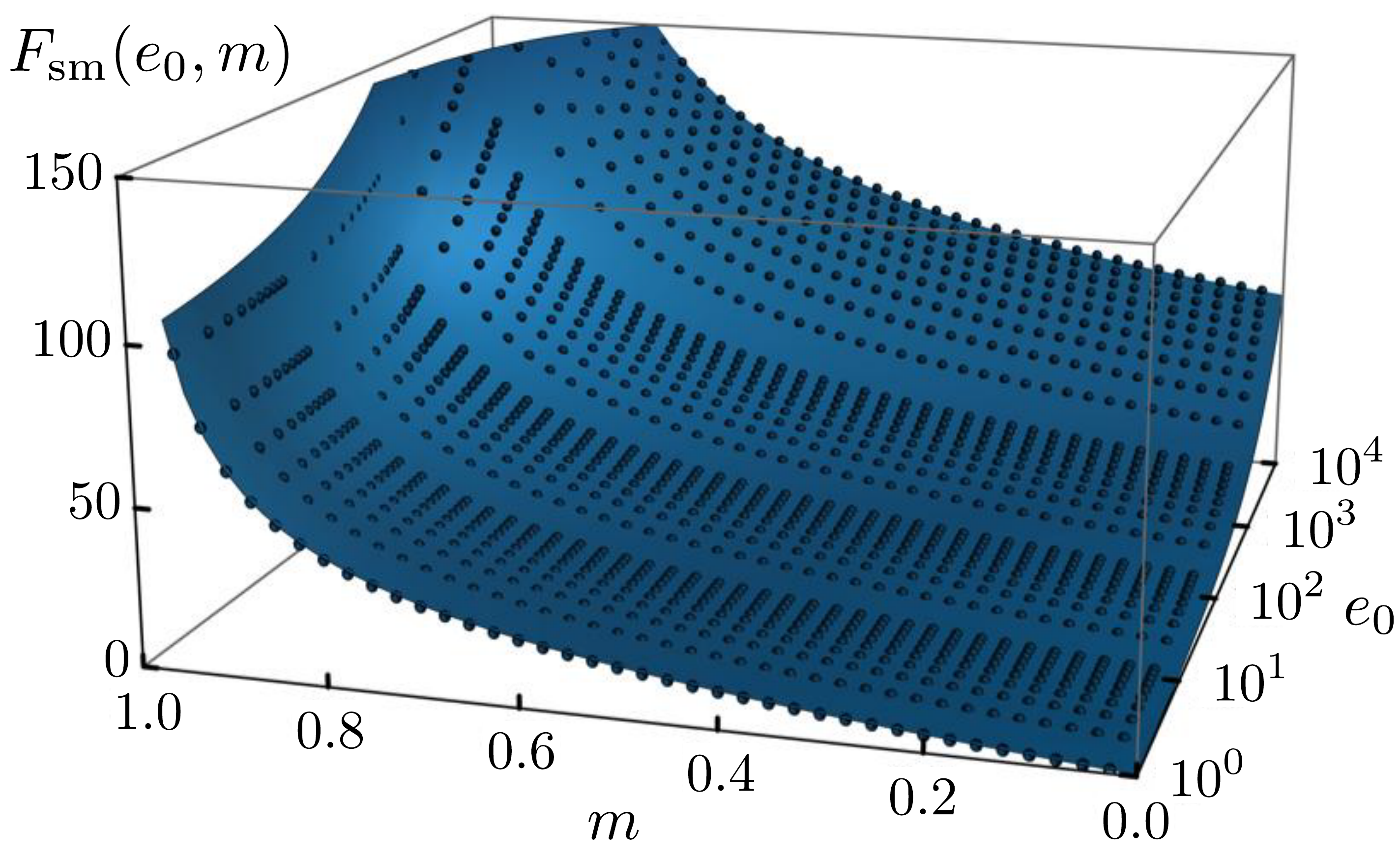}
}
\caption{\label{fig3}  The results of the numeric integration of the free energy functional for Riemann's minimal surfaces for varying values of the elliptic modulus $m$ and the cutoff energy $e_0$ are plotted along with the fit in Eq.~(\ref{energy_fit}).
}
\end{figure}

In order to calculate the free energy of Riemann's minimal surfaces, we numerically integrated $F=\int dV f_{\rm comp} \theta(e_0-f_{\rm comp}),$ where $f_{\rm comp}$ is the compression energy density $(B/2) (1-\nabla \Phi^2)^2$ and $\theta(x)$ is the Heaviside theta function, defined to be $1$ when $x>0$ and $0$ when $x<0$.  The resulting energies depend on both the elliptic modulus $m$ and the free energy cutoff $e_0$. As with the single screw dislocation, we expect the energy to scale as the square root of the cutoff, and, indeed, for fixed elliptic modulus, the energy fits $F_{m}(e_0)=A+B\sqrt{e_0},$ where the $m$-dependence of the parameters is $A(m)=a_1 m/(1-m)$ and $B(m)=b_1+b_2m /(1-m).$  
Thus, the free energy for Riemann's minimal surfaces is fit by
\begin{equation}\label{energy_fit}
F(m,e_0)=a \frac{m}{\sqrt{1-m}}+\sqrt{e_0} \Big(b+c\frac{m}{1-m}\Big),
\end{equation}
for parameter values $a=21.7622 \pm 0.0300$, $b=0.6319 \pm 0.0009 $ and $c=0.1387 \pm 0.0003.$
Though there is no justification for this functional form, the fit, shown in FIG. \ref{fig3}, is quite good.  
It would be interesting to understand the success of this fit.  Note that the constant proportional to $\sqrt{e_0}$ is close to $2/\pi \approx 0.6366$, giving the contribution from two non-interacting dislocations.  The $m$ dependence becomes a $1/r^2$ energetic interaction between the defects for large separations $r$.

\section{Discussion}

Nowhere is the ``duality'' between the two topological descriptions of Riemann's minimal surfaces more important than in the energetics.    By choosing the helicoidal description of the phase field, the divergences in the energetics lie along the defect cores.  Here the smectic layers are flat and uniformly spaced at infinity.  Conversely, a smectic containing a pore will have curvature singularities down the center of the pore and at infinity, reminiscent of the focal lines in the focal conic texture.

This simple example highlights the importance of topological decomposition of minimal surfaces in studying complex systems in nature.  Enumerating all topological defects in a surface simultaneously identifies all energetic singularities.  This technique will help explain a variety of phenomena -- in particular, triply periodic minimal surfaces which appear in bicontinuous cubic phases of systems ranging from mitochondrial membranes to the dark conglomerate phase of bent core liquid crystals to binary metallic alloys. The network of defects recovered from a topological decomposition should explain both their extraordinary stability and complex phase diagram. By tuning the interfacial preference for negative Gaussian curvature, the cubic phases continuously transform into one another following the progression Schwarz P surface to diamond surface to gyroid. Yet simple curvature considerations cannot explain this series of surfaces. Although these surfaces are topologically distinct, tracking the evolution of the lattice of their defects may elucidate the means by which these surfaces smoothly change topology \cite{Fogden:1999p91}.

\end{document}